\documentclass[a4paper]{article}
\usepackage{amssymb}
\usepackage{amsmath}
\usepackage{latexsym}
\def\setR{\mathbb{R}}

\def\setC{\mathbb{C}}
\def\calH {{\cal H}}
\def\ie {{i.e.}}

\newcommand{\sss}[1]{\scriptscriptstyle #1}
\newcommand{\bs}[1]{\boldsymbol{#1}}

\newcommand{\lbl}[1]{\label{eq: #1}}
\newcommand{\rf}[1]{\ref{eq: #1}}

\begin{document}
\begin{center}
{\huge De Sitter Waves and the Zero Curvature Limit}\\
\bigskip
{\large 
T. Garidi$^{1,2}$, 
E. Huguet$^{2,3}$, 
J. Renaud$^{1,2}$}
\end{center}
{\small
$1$ - LPTMC, Universit\'e Paris 7  Denis Diderot, boite 7020
F-75251 Paris Cedex 05, France.\\
$2$ - F\'ed\'eration de recherche APC, Universit\'e Paris 7
Denis Diderot, boite 7020\- F-75251 Paris Cedex 05, France.\\
$3$ - GEPI, Observatoire de Paris 5. pl. J. Janssen, 92195 Meudon
Cedex, France.}
\footnote{garidi@kalymnos.unige.ch} 
\footnote{eric.huguet@obspm.fr} 
\footnote{renaud@ccr.jussieu.fr}
\bigskip
\begin{abstract}
We show that a particular set of global modes for the massive
de~Sitter scalar field (the de~Sitter waves) allows to manage the
group representations and the Fourier transform in the flat
(Minkowskian) limit. This is in opposition to the usual acceptance
based on  a previous result, suggesting the appearance of negative
energy in the limit process. This method also confirms that the
Euclidean vacuum, in de~Sitter spacetime, has to be preferred as
far as one wishes to recover ordinary QFT in the flat limit.
\end{abstract}

\newpage

\section{Introduction}
A major issue in the resurgence of de Sitter (dS) space physics
motivated by inflation scenarii \cite{daniel1,daniel2},
astronomical observation \cite{perl}, dS/CFT correspondence
\cite{antoniadis,mazur,bousso,tolley}, and the study of a simple maximally
symmetric space with non vanishing curvature, concerns the status
of a ``preferred" vacuum state for the associated QFT. The absence
of a global time-like Killing vector field in de Sitter space (non
stationary) excludes the ``natural" choice characterized by the
spectrum of a Hamiltonian operator unlike the Minkowski case.
The presence of a maximal
symmetry group does not get rid of this problem:
 there exists a family of inequivalent vacua which are all invariant
under the dS group \cite{chernikov,mottola,allen1}.

Nevertheless, thanks to  this group, one can study the limit at
vanishing curvature owning to the method of group contraction
which allows to follow the unitary irreducible representations
(hereafter UIR) in that limit. It has been shown \cite{mini} that
the representations of the de Sitter group associated to the
massive scalar field, \ie\ the principal series of SO$(1,4)$,
contract (in the zero curvature limit) toward the direct sum of
two UIR's of the Poincar\'e group associated respectively to
positive and negative frequencies massive scalar fields, namely:
\begin{equation}\label{contract}
D_{\nu}\,\longrightarrow \,{\cal P}(+m)\oplus{\cal{P}}(-m)\,.
\end{equation}
This result could appear as somewhat confusing since it suggests
that the curvature is in some sense responsible for the emergence
of negative frequency modes in QFT. This is all the more
disturbing since a recent paper shows that these modes necessarily
occur in the covariant quantization of the minimally coupled
scalar field \cite{gareta1}. Since on the level of two-point
functions, the flat limit seems to work perfectly well, it has
been argued that group representation contractions were not
adapted for the study of QFT \cite{brmo}. Attempts have been made
in replacing SO$(1,4)$ by the de Sitter ``causal semi-group''
which contracts toward the Poincar\'e causal semi-group
\cite{mizony}. In view of the decisive role played by group theory
in ordinary QFT and in defining on de~Sitter space objects as mass
or spin, it is really frustrating that one cannot manage group
representation in the flat limit process. In this paper we
propose to amend this drawback.

The Euclidean vacuum has been studied before and singled out by
analyticity requirements \cite{brmo,brgamo}, flat space behavior
or further reasons listed in \cite{larsen}. Although the Euclidean
vacuum seems to be favored, it remains sensible to use the  whole
vacua family; for instance as tools in order to investigate the
effects of transplanckian physics \cite{daniel1,daniel2}. In this
paper, we reconsider the flat limit through the modes. The flat
limit for a mode is obtained by considering the latter on a domain
which is small compared to the inverse of the curvature. This
process can be applied of course at any point of spacetime with
different results. The use of ambient space formalism allows to
show in a very simple way that the Euclidean vacuum is the only
vacuum for which the flat limit yields, {\em in any point of
spacetime}, positive frequency modes.  Furthermore the use of the
de~Sitter waves shows that  the whole free QFT  tends toward the
flat theory when the curvature vanishes, including the de~Sitter
Fourier transform which becomes the ordinary Fourier transform in
the limit. Some of us will show in future works that these
de~Sitter waves are also very well adapted to group
representations and spinorial computation.

Moreover, our procedure will allow us to reconsider the
significance of the result on group contractions quoted before. In
this paper we argue that although Eq. (\ref{contract}) {\it
can} hold it does not represent the only possibility. Actually, we
show that the principal series of SO$(1,4)$ can contract toward
the positive energy representation of the Poincar\'e group, result
which is, as soon as we know, new.

The de~Sitter waves and ambient space formalism are summarized in 
Sec. 2. The flat limit is investigated in Sec. 3. 
The problem of the contraction of group representations is 
tackled in Sec. 4. Sec. 5. is devoted to some concluding remarks. 

\section{The de sitter waves}
The de Sitter space is conveniently seen as a hyperboloid embedded
in a five-dimensional time oriented Minkowski space $E_{\sss 5}$:
$$
M_{\sss H}\,=\,\{ X\in E_{\sss
5}\,|\,X^2=\eta_{\alpha\beta}X^\alpha X^\beta\,=\,-H^{-2}\}\,,$$
where $\eta^{\alpha\beta} =\mbox{diag}(1,-1,-1,-1,-1)$. The
(pseudo-) sphere $M_{\sss H}$ is obviously invariant under
 O$(1,4)$. We only consider the
connected component of the identity SO$_{\rm o}(1,4)$, the
so-called de~Sitter group. We are in particular interested in the
flat limit (\ie~$H \to 0$) of the massive
scalar free quantum field  and the behavior of the group
representation in this limit.

The free massive scalar field on this spacetime is, in the Wigner
sense, an {\em elementary system} whose associated unitary
irreducible representation belongs to the principal series of
representations of SO$_{o}$(1,4). This UIR is characterized by
 the eigenvalue $\nu^{2}+9/4$
 of the Casimir operator $Q_{0}$  which is linked to
  the Laplace-Beltrami operator on
$M_{\sss H}$ through $-H^{2}Q_{0}\equiv\Box_{\sss H}$ \cite{dix}.
 The contraction of that UIR has already been studied in a
group theoretical context \cite{mini}. The result is usually
written in the following way: the massive representations contract
toward the direct sum of the positive energy and negative energy
representations of the Poincar\'e group. We emphasize that this
result has been achieved on a purely group theoretical level
through an ad hoc process of contraction. Although it is from this
point of view remarkable that the irreducible representation {\em
can contract} toward a reducible representation,  there is no
uniqueness in this choice of contraction procedure. In the
framework of QFT this result played a rather misleading role in
order to understand field theory on de~Sitter background from our
Minkowskian point of view. Actually, we will see that the negative
energy plane waves do not appear when the curvature vanishes as
soon as the Euclidean vacuum has been chosen.

In \cite{brmo,brgamo}, the authors use a set of global modes, the
de~Sitter waves, solutions of the de~Sitter Klein-Gordon equation,
which are the formal analogue of the plane waves in Minkowski
spacetime. We will see that these modes reduce to the usual plane
waves when the curvature tends to zero as far as their analyticity
domain has been conveniently chosen.

Let ${\cal C}^{+}\,=\{\xi\in E_{5};\;
 \xi^{2}=0,\, \xi^{0}>0\}$ be  the null upper cone of
$E_{\sss 5}$. The multivalued functions defined on dS spacetime
by:
\begin{equation}
X\mapsto\left(HX \cdot \xi\right)^s, \,\xi\in{\cal C}^{+},\,
X\cdot\xi\,\neq\,0,\, s\in\setC,\label{eq:mode}
\end{equation}
are solutions of the de~Sitter Klein-Gordon equation
$(\square_{\sss H}+m^2+12 H^2\zeta)\phi=0$, where $\zeta$ is a
positive gravitational coupling with the de Sitter background and
$$s=-\frac{3}{2}- i\nu\ \;\;\mbox{where}\;\;
\nu=\frac{1}{2}\sqrt{4m^{2}H^{-2}+48\,\zeta - 9}\;\in \setR \,,
$$ corresponds  to the principal series of UIR (massive
case). The expression defined by Eq. ($\ref{eq:mode}$) is, {\em as
a function of $\xi$}, homogeneous with degree $s$ on ${\cal
C}^{+}$ and  thus is entirely determined by specifying its values
on a well chosen three dimensional submanifold (the so-called
orbital basis) $\gamma$ of ${\cal C}^{+}$. These dS waves, as
functions on de Sitter spacetime, are only locally defined because
they are singular on specific lower dimensional subsets of
$M_{{\sss H}}$ and multivalued since $\left(HX \cdot \xi\right)$
can be negative. In order to get a singlevalued global definition,
they have to be viewed as distributions which are boundary values
of analytic continuations to suitable domains in the complexified
de Sitter space $M_{\sss H}^{(c)}$:
\begin{eqnarray*}
M_{\sss H}^{(c)}&=&\{Z=X+iY\in E_{5} + i E_{5}; \eta_{\alpha
\beta} Z^\alpha Z^\beta=-H^{-2}\}.
\end{eqnarray*}
 The minimal domains of analyticity which yield
single-valued functions on de~Sitter spacetime are the forward and
backward tubes of $M_{\sss H}^{(c)}$: ${\cal T}^\pm=T^\pm\cap
M_{\sss H}^{(c)}$, where $T^\pm=E_{5}-iV^\pm$ and $V^\pm=\{ X\in
E_{5};\;\; X^0\stackrel{>}{<}\sqrt{\parallel \bs{X} \parallel^2 +
(X^4)^2} \}. $ Details are given in \cite{brmo}.

When $Z$ varies in ${\cal T}^+$
 and $\xi$ lies in the positive cone
${\cal C}^+$, the  functions given in Eq. (\ref{eq:mode}) are
globally well defined since the imaginary part of $(Z.\xi)$ is non
positive. We define the de~Sitter waves $\phi_{\xi}(X)$ as the
boundary value of the analytic continuation to the future tube of
Eq. (\ref{eq:mode}):
\begin{eqnarray}\phi_{\xi}(X)&\equiv& c_{\nu}\mbox{bv}(HZ\cdot\xi)^{s}
\, = \, c_{\nu} \bigl [\theta(H X\cdot \xi) \nonumber \\ &+&
\theta(-H X\cdot\xi)\,e^{-i\pi s }\bigr ] |H
X\cdot\xi|^{s},\lbl{phi}
\end{eqnarray}
where $\theta$ is the Heaviside function. The real valued constant
$c_{\nu}$ is determined by imposing the Hadamard condition on the
two-point function. This choice of modes corresponds to the
Euclidean vacuum. In terms of de Sitter  waves, the two-point
function reads \cite{brgamo}:
\begin{equation*}
W(z,z')={c_{\nu}}^2\int_{\gamma}(HZ\cdot\xi)^{s}(HZ'\cdot\xi)^{s^*}
d\sigma_{\gamma}(\xi)\,,
\end{equation*}
where  $Z\in{\cal T}^+$ and $Z'\in{\cal T}^-$. The measure
$d\sigma_{\gamma}(\xi)$ on the orbital basis $\gamma$ is chosen to
be  $m^2$ times the natural one  induced from the $\setR^{5}$
Lebesgue measure. The calculation, similar to that of
\cite{brgamo} yields:
\begin{equation*}
c_{\nu}=\sqrt{\frac{H^{2}\left(\nu^2+1/4\right)}
{2(2\pi)^3(1+e^{-2\pi\nu})m^{2}}}.
\end{equation*}

\section{The flat limit of de sitter waves}
Hereafter, we investigate the behavior of the mode $\phi_{\xi}(X)$
under vanishing curvature. We consider a region around any point
$X_{\sss A}$ in which all the distances are small compared to
$H^{-1}$. With this assumption, we will prove that
\begin{eqnarray}
\lim_{H\rightarrow 0}\phi_{\xi}(X) &=&\frac{1}{\sqrt{2(2\pi)^3}}
\exp(-ikx)\, ,\,\mbox{for} \,X_{\sss A} \cdot \xi\,>\,0,\nonumber\\
\lim_{H\rightarrow 0}\phi_{\xi}(X) &=& 0,\,\mbox{for} \,X_{\sss A}
\cdot \xi\,<\,0 .\label{eq:lim}
\end{eqnarray}
In other words, these modes do not generate negative frequency
modes in the flat limit, whatever the point around which the limit
is computed.

Due to the homogeneity of the de~Sitter space under the de~Sitter
group action, one can choose a system of coordinates
such that
 $X_{\sss A}^{4}=H^{-1}$ and $X_{\sss A}^\mu=0$. In the
neighborhood of this point, for $H\to 0$, the de~Sitter spacetime
meets its tangent plane (the four dimensional  Minkowski
spacetime), and the coordinates $X$ of this neighborhood read:
\begin{equation}\lbl{coord}\left\{\begin{array}{rcl}
X^\mu&=& x^\mu+o(H)\,\\
X^4&=& H^{-1}+o(1)\,.
\end{array} \right.\end{equation}
For $s\sim-\frac{3}{2}-imH^{-1}$,  $\exp(-i\pi s)\to 0$ and one
obtains:
$$\lim_{H\rightarrow 0}\phi_{\xi}(X)=\lim_{H\rightarrow 0}
c_{\nu}\,\theta(H X\cdot \xi)|H X\cdot\xi|^{s}. $$ The Heaviside
function yields $\xi^{4}<0$ since $H X\cdot \xi\simeq-\xi^{4}$ and
finally, for small  $H$:
$$\phi_{\xi}(X)\simeq
\frac{\,|\xi_{4}|^{s}}{\sqrt{2(2\pi)^3}} \left(1+\frac{H \xi_{\mu}
x^{\mu}}{|\xi_{4}|}\right)^{-\frac{3}{2}-imH^{-1}}\theta\left(-\xi^{4}\right)
\;.$$
This limit exists only for $|\xi_{4}|=1$. As a consequence, we use
the orbital  basis $\gamma={\cal C}_{1}\cup{\cal
C}_{2}\,\lbl{ob}$, where ${\cal C}_1,{\cal  C}_2$ are defined by:
$$\xi=(\frac{\omega_{\boldsymbol{k}}}{m},\,\frac{\boldsymbol{k}}{m},\,-1)
\in{\cal
C}_{1},\quad\;\xi=(\frac{\omega_{\boldsymbol{k}}}{m},\,
\frac{\boldsymbol{k}}{m},\,1)\in{\cal
C}_{2},
$$ with $\omega_{\boldsymbol{k}}~=~\sqrt{\boldsymbol{k}^2~+~m^2}$. 
Note that the induced
measure on $\gamma$ is $d\boldsymbol{k}/(m^{2}\omega_{\boldsymbol{k}})$ and
therefore $d\sigma_{\gamma}(\xi)=d\boldsymbol{k}/\omega_{\boldsymbol{k}}$. We
finally obtain Eqs. ($\ref{eq:lim}$) according to whether $\xi$ belongs
to ${\cal C}_{1}$ or ${\cal C}_{2}$  i.e.,  $X_{\sss A}\cdot\xi$
positive or negative.

Thus, due to the analyticity condition at the origin of the
$\exp(-i\pi s)$ term, the negative energy modes are
(exponentially) suppressed whereas the positive energy modes give
the Minkowskian on-shell modes corresponding to a particle of mass
$m$.

{\em We insist on the fact that the result leads to positive
frequency plane waves whatever the point $X_{\sss A}$ we choose.}
This choice of modes, which corresponds to the Euclidean vacuum,
is the only one having this property. Any Bogoliubov
transformation on these modes leads to the appearance of conjugate
modes $\phi_{\xi}^{*}$ whose flat limit at some point $X_{\sss B}$
is a negative frequency mode as soon as $B$ has been chosen in such a way
that $X_{\sss B} \cdot \xi < 0$.

As a consequence, any vacuum different from the Euclidean vacuum
would lead to physically unacceptable Minkowskian QFT. The
Euclidean vacuum has therefore to be preferred with respect to the
flat limit criterion.

The de~Sitter waves allow to define a de~Sitter Fourier transform
which becomes the ordinary Fourier transform in the flat limit. In
fact, one can realize the de~Sitter one particle sector ${\cal
H}_{\sss H}$ as distributions on spacetime through this de~Sitter
Fourier transform: any $\psi\in{\cal H}_{\sss H}$ can be written
as:
\begin{equation}
\psi(X)\,=\,\int_{\xi \in \gamma}
\phi_{_\xi}(X)\,\tilde{\psi}(\xi)\,
d\sigma_{\gamma}(\xi),\;\tilde\psi\in
L^{2}(\gamma,d\sigma_{\gamma}),
\end{equation}
see \cite{brmo} for details. Let $X_{\sss A}$ be a point of
de~Sitter spacetime in the neighborhood of which we will proceed
to the flat limit. The space ${\cal H}_{\sss H}$ can then be
decomposed into ${\cal H}_{\sss H}={\cal H}_{\sss H}^{1}\oplus
{\cal H}_{\sss H}^{2}$ using the decomposition of the orbital
basis:
\begin{eqnarray} \label{eq: bo}
\psi(X) &=&\int_{\xi \in {\cal C}_{1}} \phi_{_\xi}(X)\,
\tilde{\psi}(\xi)\, d\sigma_{\gamma}(\xi) \nonumber \\
&+&\, \int_{\xi \in {\cal
C}_{2}} \phi_{_\xi}(X)\,\tilde{\psi}(\xi)\, d\sigma_{\gamma}(\xi).
\end{eqnarray}
In the limit of null curvature, the second integral of the above
expression vanishes and only the positive frequency remains:
$$\lim_{H\to 0}\psi(x)=\int
\frac{ e^{-ikx}}{\sqrt{2(2\pi)^3}}\,
\tilde{\psi}(k)\,\frac{d\boldsymbol{k}}{\omega_{\boldsymbol{k}}}.$$

As a consequence, the ordinary Fourier transform is the flat limit
of the de~Sitter Fourier transform. Once again, one can see the
significance of de~Sitter waves which play in de~Sitter space the
role of plane waves in Minkowski space, including a good behavior
with respect to the de~Sitter group:  one can see easily using
$\phi_\xi (g^{-1}\! X) = \phi_{g\hspace{-0.009cm}\xi}(X)$ that
each space ${\cal H}_{\sss H}^{i}$ is invariant under the subgroup
generated by the $M_{ab}$ with $0\leq a<b\leq3$ (see appendix).
This subgroup, isomorphic to SO$_{o}(1,3)$, is the stabilizer of
$X_{\sss A}$.

\section{Group contraction}

The Minkowski spacetime is the flat limit of the de Sitter
spacetime with respect to  all the objects of QFT. In order to
emphasize this fact and clarify the link between our approach and
that of \cite{mini} we will present the concept of contractions in
a slightly different manner from the usual presentation.

 Let us consider a family of representations
   $U^{\sss H}$ of a group $G$ into  some spaces
${\cal H}_{\sss H}$ and a representation $U$ of a group $G'$ into
a space $\cal H$. One wants to give a precise meaning  to the
assertion $U^{\sss H} \to U$ for $H\to 0$ (one says that the
representations $U^{\sss H}$ contract toward $U$).

First, we must have a bijection $G\stackrel{i}{\rightarrow} G'$
(which is not an homomorphism) conveying the ``similarity''
between the two groups.  Second we need a space, equipped with a
topology, in which all the representations are written. This is
obtained by writing an injective map $A_{\sss H}$ from ${\cal
H}_{\sss H}$ to $E\supset \cal H$ where $E$ is a topological space
containing $\cal H$ in such a way that for any $\phi\in {\cal
H}_{\sss H}$ the limit ${\displaystyle\lim_{{\sss H}\to 0}A_{\sss
H}\phi=h}$ exists in $E$ and belongs to $\cal H$.

We say that the representations $U^{\sss H}$ contract toward $U$
if:
\begin{equation}
\forall \psi \in {\cal H}_{\sss H}, \;
\lim_{H\to 0} A_{\sss H}U_{g}^{\sss H}\psi=U_{g'}h=
U_{g'}\lim_{H\to 0}A_{\sss H}\psi,\lbl{cont}\end{equation} where
$g'$ is the element of $G'$ identified with $g\in G$ by means of
$i$.

Let us now return to de Sitter context. For $x$ in Minkowski space
we define $X$ in de~Sitter space through Eq. (\rf{coord}). We then
can define $A_{\sss H}$: for the de~Sitter waves $\phi_{\xi}$ we
define $A_{\sss H}\phi_{\xi}$ as a function on Minkowski spacetime
through:
$$\left(A_{\sss H}\phi_{\xi}\right)(x)=\phi_{\xi}(X).$$
This definition  extends linearly to ${\cal H}_{\sss H}$ through
Eq. (\ref{eq: bo}). Then in view of Eq. (\ref{eq:lim}) we have, at
least in a weak sense, for any $\psi$ in ${\cal H}_{\sss H}$:
$$\lim_{H\to 0}A_{\sss H}\psi= f,$$
where $f$ is a positive frequency wave packet on Minkowski
spacetime.

We now turn to the representations. The de~Sitter and Poincar\'e groups
 are identified as explained in the appendix. Consider:
\begin{eqnarray*}
g&=&\exp\biggl(\sum_{ab}\alpha_{ab}M_{ab}\biggr)\stackrel{i}{\mapsto}
 g'\,=\,\exp\biggl(\sum_{j}\alpha_{0j}B_{j} \\
&+&\sum_{ij}\alpha_{ij}R_{ij}
+\sum_{\mu}\alpha_{\mu 4}T_{\mu}\biggr).
\end{eqnarray*}
The representation of the de Sitter group is defined by:
$$U^{\sss H}_{g}\psi(X)=\psi(g^{-1}_{\sss H}\! X),$$
where $g_{\sss H}$ is the $5\times5$ matrix defined by
$$g_{\sss H}=\exp(\sum_{a<b<4}\alpha_{ab}M_{ab}
+H\sum_{\mu}\alpha_{\mu4}M_{\mu4}),$$ and the representation of
the Poincar\'e group is defined by:
$$U_{g'}\psi(x)=\psi({g'}^{-1}\! x).$$
One can easily see that for $\xi\,\in{\cal C}_{1}$:
\begin{equation}
\left(H (g^{-1}_{\sss H}\!X)\cdot\xi\right)=\left(H (g'^{-1}\!x)\cdot
k+\,o(H)+1\right)\,.
\end{equation}
Then Eq. (\rf{cont}) follows and the principal series of
representations of the de~Sitter group contract toward the
positive energy representation of the Poincar\'e group. Once
again, no negative energy can appear in this process. Nevertheless
this is not in contradiction with \cite{mini} for which this
series can contract toward another representation. In fact, in our
context, the result of \cite{mini} can be recovered by modifying 
$A_{\sss H}$ in the following way. One can define $\tilde A_{\sss H}$ 
by $\tilde
A_{\sss H}=A_{\sss H}$ on ${\calH}_{\sss H}^{1}$ and $\tilde
A_{\sss H}=\exp(+i\pi s) A_{\sss H}$ on ${\calH}_{\sss H}^{2}$.
With this operator, one obtains the result of \cite{mini} because
the artificial exponential term cancels the natural one which is
present in the definition of $\phi_{\xi}$ thanks to the property
of analyticity.

\medskip
\section{Conclusions}
Recently, several papers summarized the theory of
irreducible unitary representations of the de Sitter group. 

A result, commonly quoted in these summaries, suggests that the 
appearance of negative energies for a Minkowskian observer is an
unavoidable consequence of group theory. For this reason some
authors claimed that the contraction procedure of group representations 
was not suitable in order to investigate the flat limit of dS-QFT.

We have shown that this is inexact. We also
conclude that the Euclidean vacuum as to be preferred as
far as one wishes to recover ordinary QFT in the flat limit. 

To that
end we used the formalism of de~Sitter waves which turned out to be a very 
convenient tool, possibly as useful in de~Sitter space as the 
plane waves in Minkowski space.  

\subsection*{Acknowledgments}

The authors thank J. Bros, J-P. Gazeau, M. Lachi\`eze-Rey and U.
Moschella for valuables discussions and helpful criticisms.

\section*{Appendix: {\rm Identification of de~Sitter and Poincar\'e groups}}

We begin with Lie algebras.
Let $\Delta_{a,b}$ be the $5\times 5$ matrices whose entries $a_{nm}$
are defined by $a_{nm}=\delta_{na}\delta_{mb}$.

The following matrices are a basis of the Lie algebra so(1,4) of the
Lie group SO$_{o}$(1,4).
\begin{eqnarray*}
M_{0b}&=&\Delta_{0b}+\Delta_{b0}\quad\mbox{ for } b=1,2,3,4\\
M_{ab}&=&\Delta_{ab}-\Delta_{ba}\quad\mbox{ for } 0<a<b\leq4.\\
\end{eqnarray*}
The following matrices are a basis of the Lie algebra p(1,3) of the
Poincar\'e  group.
\begin{eqnarray*}
B_{j}&=&\Delta_{0j}+\Delta_{j0}\quad\mbox{ for } j=1,2,3\\
R_{ij}&=&\Delta_{ij}-\Delta_{ji}\quad\mbox{ for } 0<i<j\leq3\\
T_{\mu}&=&\Delta_{\mu4}\quad\mbox{ for } \mu=0,1,2,3.\\
\end{eqnarray*}
The identification between the two Lie algebras is obtained through :
\begin{eqnarray*}
M_{0j}&\simeq&B_{j}\quad\mbox{ for } j=1,2,3\\
M_{ij}&\simeq&R_{ij}\quad\mbox{ for } 0<i<j\leq3\\
M_{\mu4}&\simeq&T_{\mu}\quad\mbox{ for } \mu=0,1,2,3.\\
\end{eqnarray*}
The identification between the groups follows, using the exponential
application.

\end{document}